\documentstyle[psfig,preprint,tighten,floats,aps]{revtex}
\def\B{{\cal B}}

\def\P{{\cal P}}
\def\ket#1{{| #1 \rangle}}
\def\bup{B^{\downarrow}}
\def\bdown{B^{\uparrow}}

\begin{document}
\draft
\title{Quantum Geometry and \\Thermal Radiation from Black Holes}
\author{Kirill V. Krasnov\thanks{E-mail address: krasnov@phys.psu.edu}}
\address{Center for Gravitational Physics and Geometry, \\
The Pennsylvania State University, PA 16802, USA.} 
\maketitle

\begin{abstract}
A quantum mechanical description of black hole states proposed recently
within non-perturbative quantum gravity is used to study the 
emission and absorption spectra of quantum black holes. We
assume that the probability distribution of states of the quantum black
hole is given by the ``area'' canonical ensemble, in which the horizon area
is used instead of energy, and use Fermi's golden rule to find
the line intensities. For a non-rotating
black hole, we study the absorption and emission of s-waves 
considering a special set of emission lines. To find the line
intensities we use an analogy between a microscopic state of
the black hole and a state of the gas of atoms.
\end{abstract}

\section{Introduction}

Progress has recently been made within non-perturbative
quantum gravity in understanding how one can 
describe states of black holes quantum mechanically \cite{K,ABCK,ACK,ABK}. 
In this description the black hole horizon  
is treated as an interior spacetime boundary where certain boundary 
conditions, motivated 
by the geometrical properties of black hole solutions, are imposed.
The standard techniques of loop quantum gravity
can then be applied, with appropriate modifications required
by the presence of the boundary. The degrees of freedom 
associated with the black hole  then
turn out to be described by Chern-Simons theory on the boundary
(see \cite{K,ABCK}).

In the present paper we use this description to 
study emission and absorption processes by quantum black holes. 
At the outset, we wish to point out our analysis is not complete;
at some points we use heuristic arguments and make certain
plausible assumptions. However, to some extent our
assumptions are supported by the final results.

The description of black hole states proposed within non-perturbative
quantum gravity realizes in a rather explicit fashion an
old heuristic idea that dates back to early works of Bekenstein (for a 
recent reference see \cite{BM}): a quantum black hole is seen
as an `atom' with
a lot of internal microstates. A detailed description of
these microstates will be given in the following section. However,
to motivate certain constructions that follow, let us present
a simple heuristic description already at this stage. In non-perturbative
quantum gravity spatial quantum geometry, i.e., the quantum geometry
``at a given time'' is described by the so-called 
cylindrical states. A cylindrical quantum state is labelled by
a graph in space. The edges of this graph have a simple heuristic
interpretation of the flux lines of the gravitational field: one
can think of the gravitational field (metric tensor) as being
concentrated along these edges, similarly as one can think of
the usual electric field of Maxwell theory as being concentrated
along Faraday's lines of electric force. A black hole 
or, more precisely, the black hole horizon ``at a given time''
is described as a two-surface that may
be pierced by the gravitational flux lines. The flux lines 
excite certain surface degrees of freedom, which
are responsible for the black hole entropy (see Fig. 1a).

A black hole state is specified by giving a configuration
of the flux lines piercing the horizon, and by describing a quantum
state of the surface degrees of freedom. A very convenient basis of
quantum states is given by eigenstates of the operator measuring
the horizon area. For such an eigenstate each
flux line piercing the horizon is labelled by a quantum number: spin $j$.
The area comes from intersections of the flux lines with
the surface: each flux line intersecting the surface
contributes an area proportional to $\sqrt{j(j+1)}$ (see (\ref{qarea})).
Let us denote any one of the area eigenstates by 
$\ket{\Gamma}$. States $\ket{\Gamma}$
are going to play the key role in what follows.
Consider a quantum process in which 
the black hole jumps from a state $\ket{\Gamma}$ to state $\ket{\Gamma'}$,
such that the horizon area changes. This, for example, can
be a process in which one of the flux lines piercing
the horizon breaks, with one of the ends falling into
the black hole and the other escaping to infinity
(see Fig. 1b). This is an example of the emission process;
the two ends of the flux line can be thought of as the
two particle anti-particle quanta in Hawking's original
picture \cite{Hawk} of the black hole evaporation. 

\begin{figure}
\centerline{
\hbox{\psfig{figure=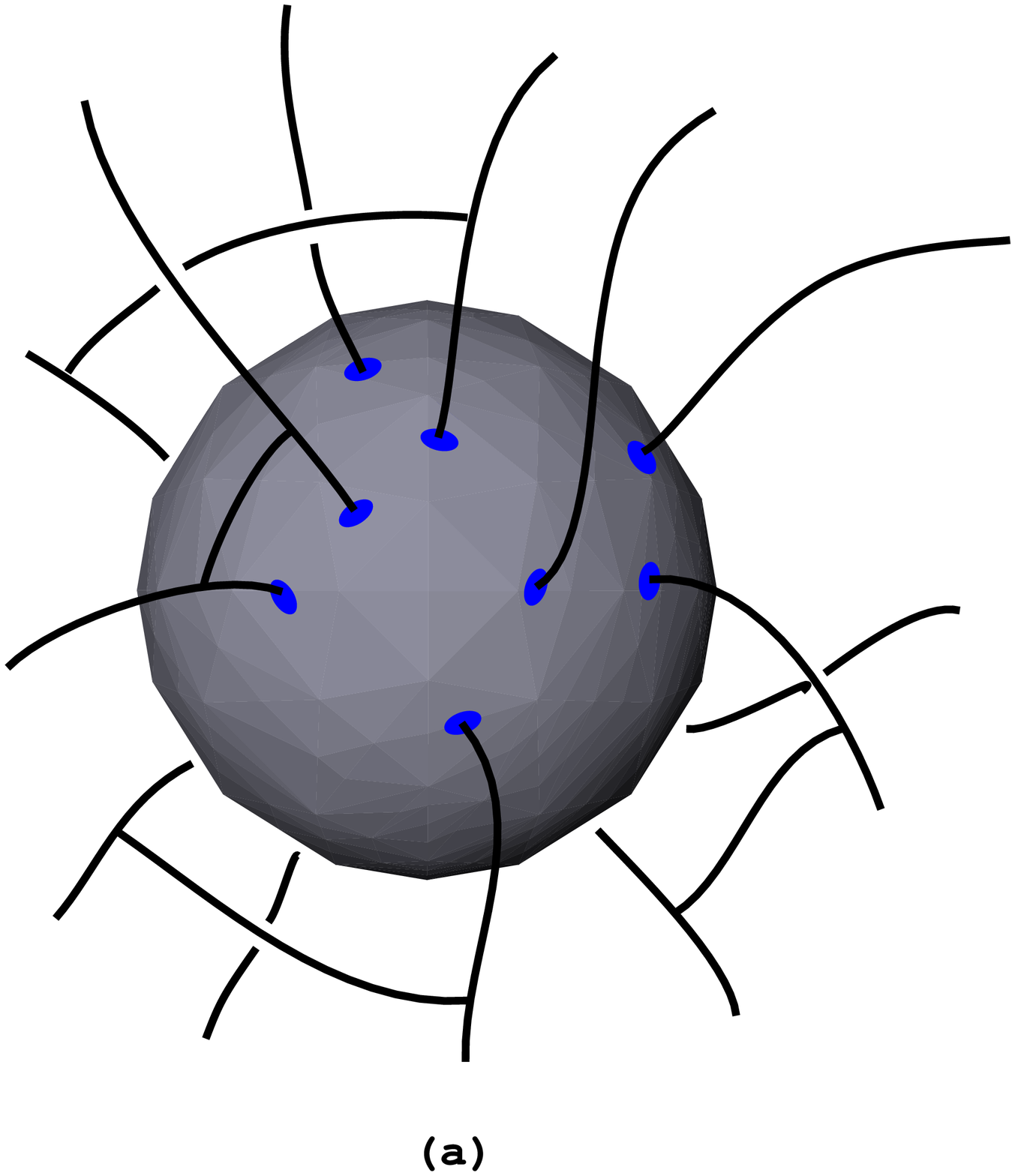,height=2in}}
\hskip0.5in
\hbox{\psfig{figure=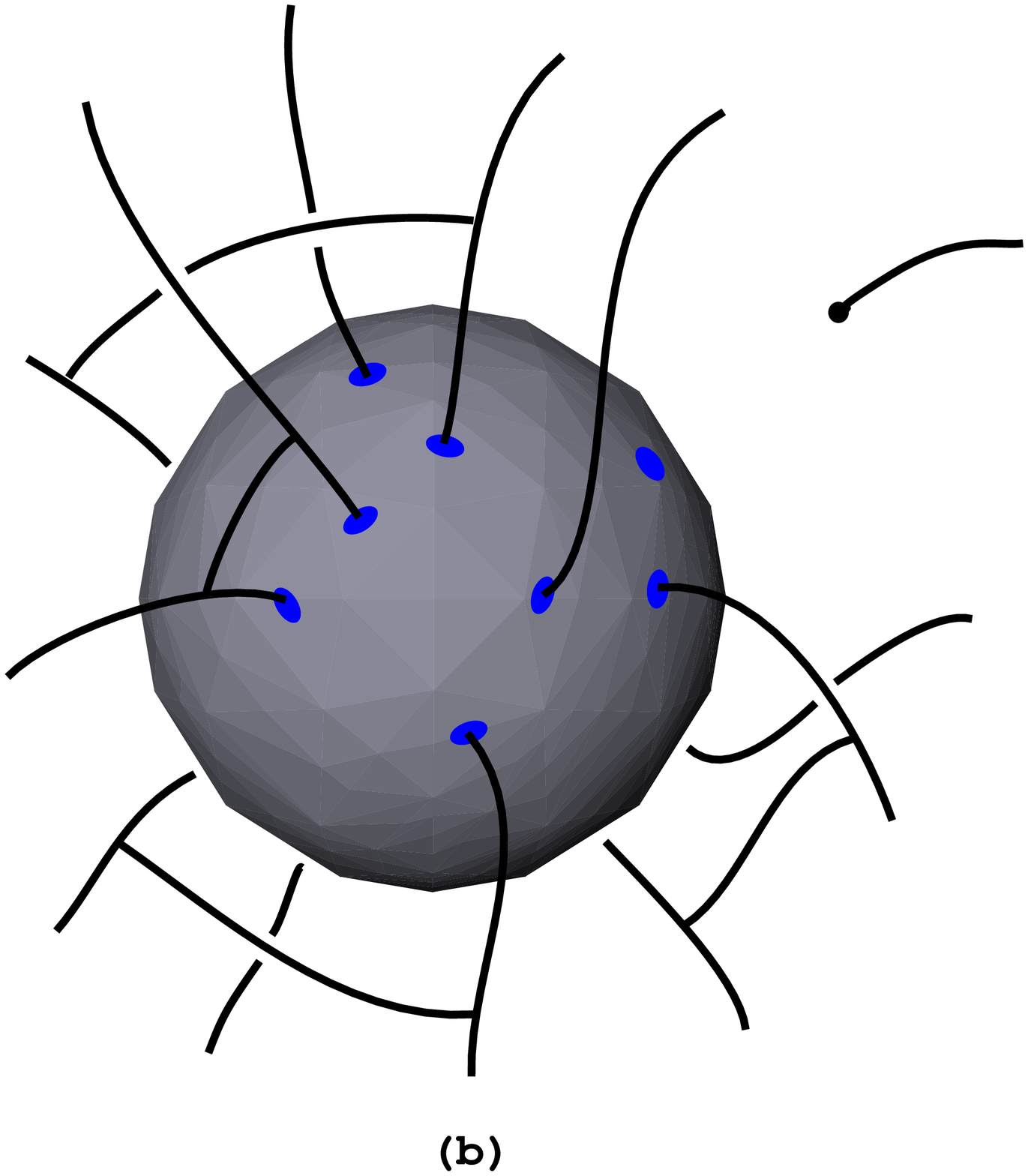,height=2in}}
}
\bigskip
\caption{(a) The flux lines of gravitational field pierce the black hole
horizon and excite curvature degrees of freedom on the surface. 
The curvature degrees of freedom, described by Chern-Simons theory, 
account for the black hole entropy in our approach; (b) An example
of the emission process in which one of the flux lines breaks,
with one of the ends falling into the black hole and the 
other escaping to infinity.} 
\end{figure}

It is not hard to find the energy of a particle that
is emitted. Let us denote by $A(M)$ the dependence of 
the horizon area of the quantum black hole
on its mass $M$ (for simplicity we consider a
non-rotating uncharged black hole), and by $\Delta M_{\Gamma\Gamma'}$ 
the change of the black hole mass in the emission process.
The quantity $\Delta M_{\Gamma\Gamma'}$ should also be equal to the energy 
of the particle emitted when it reaches infinity.
Then for small area changes $\Delta A_{\Gamma\Gamma'}$  
the energy of the particle emitted is equal to 
\begin{equation} 
\Delta M_{\Gamma\Gamma'} = \left({dA\over dM}\right)^{-1}
\,\Delta A_{\Gamma\Gamma'}.
\label{rel}
\end{equation}
For simplicity let us consider the case when 
the particle emitted is massless. Then its 
energy $\Delta M_{\Gamma\Gamma'}$  is 
related with its frequency $\omega_{\Gamma\Gamma'}$ via Planck's formula 
$\Delta M_{\Gamma\Gamma'}=\hbar\omega_{\Gamma\Gamma'}$. 
One can also consider
the case when the particle emitted carries an angular momentum.
Then the angular momentum of the particle
is equal to the angular momentum lost by the black hole,
and the relation (\ref{rel}) must be modified by replacing
the derivative with respect to $M$ by the partial derivative
with the black hole angular momentum fixed. Similarly
one can consider the emission of charged particles.

As we already mentioned, the spectrum $A[\Gamma]$ of the area is
purely discrete in our quantum theory. Thus, the black hole
spectrum is discrete: possible transitions 
$\Gamma \to \Gamma'$ correspond to discrete 
emission lines in the spectrum. However, this by itself
does not imply that the black hole spectrum is 
essentially different from the thermal spectrum predicted by
the semiclassical considerations. Indeed, as is argued
in \cite{Makela}, quantum fluctuations may smear
out the discrete lines in the emission spectrum, which
would result in an effectively continuous spectrum. 
Although the non-commutativity of certain geometrical
operators suggests \cite{Note} that this is indeed 
the case in the theory under consideration, this issue
is far from being settled. Thus, in this paper, we
simply try to find the properties of the discrete
emission spectrum, leaving the problem of determining
whether the spectrum indeed becomes effectively 
continuous to further research.

To determine the line intensities and, therefore, the form of the 
emission spectrum, 
we, motivated by the ``atomic'' picture of the black hole,
use Fermi's golden rule. According to this rule,
the probability of a transition $\Gamma \to \Gamma'$ with 
a quantum of radiation being emitted is given by
\begin{equation}
W_{\Gamma\to\Gamma'}={2\pi\over\hbar}\,|V_{\Gamma\Gamma'}|^2\,
\delta(\omega-\omega_{\Gamma\Gamma'}){\omega^2 d\omega 
d\Omega\over (2\pi\hbar)^3},
\label{fermi}
\end{equation}
where $V_{\Gamma\Gamma'}$ is the matrix element of 
the part of the Hamiltonian of
the system that is responsible for the transition. 
For brevity,
we suppress the dependence of this matrix element on initial
and final states of the quantum field describing the radiation.
Quantity $d\Omega$ in (\ref{fermi})
is the element of the solid angle in the direction of the quantum emitted, 
and $\omega_{\Gamma\Gamma'}$ is the frequency of the quantum. 
We put $c=1$ throughout the paper. The total energy $dI$ 
emitted by the system per unit time in transitions of this particular 
type can be obtained
by multiplying the above probability by $\hbar\omega$ and
by the probability $p(\Gamma)$ to find the system in the
initial state $\ket{\Gamma}$
\begin{equation}
dI(\Gamma \to \Gamma') = \hbar\omega\,p(\Gamma)\,{2\pi\over\hbar}\,
|V_{\Gamma\Gamma'}|^2\,\delta(\omega-\omega_{\Gamma\Gamma'})
{\omega^2 d\omega d\Omega\over (2\pi\hbar)^3}
\label{fermi-2}
\end{equation}

Thus, to find the emission spectrum, one
has to know the probability distribution $p(\Gamma)$ of the 
black hole over states $\ket{\Gamma}$, and the matrix
elements $|V_{\Gamma\Gamma'}|$ of the Hamiltonian 
responsible for transitions. However, as we know
from the numerous examples from statistical mechanics, 
it usually happens that the general form of the spectrum is
determined by the probability distribution over states
and by such qualitative aspects of the dynamics
as selection rules for quantum transitions.
Usually only the fine details of the spectrum depend
on the precise form of the matrix elements of the Hamiltonian.
In this paper we try to learn as much as possible about
the black hole emission spectrum without specifying the
details of the quantum dynamics, using just the information 
provided by the kinematics of the theory. 
As we shall see, it is indeed
possible to determine the general form of the spectrum just
by knowing the probability distribution $p(\Gamma)$
and by analyzing certain kinematical selection rules.

As the statistical distribution 
$p(\Gamma)$ we use the one given by the ``area'' canonical ensemble
\cite{GeomEntr}. Namely, we take the probability $p(\Gamma)$ 
to find the system in a quantum state $\Gamma$ to be
$p(\Gamma)\propto\exp{(-\alpha A[\Gamma])}$, where $A[\Gamma]$ is
the ``quantum'' area of the black hole in the state $\Gamma$,
and $\alpha$ is a real positive parameter, playing the
role of the quantity conjugate to the area.  Some properties
of this statistical ensemble are reviewed in the Appendix A. 
In particular, as we discuss in the appendix, 
the problem of calculation of the black hole entropy $S$ 
can be reduced to the problem of calculation of the
entropy in this statistical ensemble. It turns out 
that large black holes correspond to the values of $\alpha$
close to certain critical value $\tilde\alpha$. The same 
critical value $\tilde\alpha$ determines the proportionality
coefficient between the entropy and the area: $S=\tilde\alpha A$.
Thus, for large black holes, the relevant values of the parameter
$\alpha$ are the ones close to the critical value $\tilde\alpha$,
and the probability distribution of states is given by:
\begin{equation}\label{distr}
p(\Gamma) \propto e^{-\tilde\alpha A[\Gamma]},
\end{equation}
where, as we have said above, $\tilde\alpha$ coincides with 
the proportionality coefficient between the 
entropy and the area $S=\tilde\alpha A$. It is clear that this
probability distribution is quite natural in the 
context of black holes. Indeed, (\ref{distr}) simply states
that the probability to find the black hole in 
a state $\Gamma$ is proportional to $\exp{(-S)}$, where $S$
is the entropy of the black hole of the horizon area $A[\Gamma]$.
Thus, if one replaces $S$ by the logarithm of the 
corresponding number $N$ of available states, then 
(\ref{distr}) becomes the ordinary microcanonical ensemble
distribution:
$p(\Gamma)\propto 1/N$, in other words, all states $\Gamma$ 
whose area is close to $A$ have equal probability $1/N$,
where $N$ is the total number of such states.

The probability distribution (\ref{distr}) is quite
reminiscent of the usual canonical ensemble. Thus, 
as the standard statistical mechanical argument 
guarantees that the emission spectrum of any system 
described by the canonical ensemble is thermal, one
could expect the same standard argument to guarantee 
the thermal character of the emission spectrum in our case.
Unfortunately, one cannot refer directly to this argument
because the area in used in (\ref{distr}) instead of energy.
A minor modification of the argument is necessary in order
to make it applicable to the area canonical ensemble. 
We give such a modified argument in the Appendix B.

Thus, as we show in the appendix, our 
choice of the probability distribution
guarantees that the mean number $n_{\omega lm}$ of 
emitted quanta in the mode of frequency $\omega$ and angular 
momentum quantum numbers $l,m$ is given by
\begin{equation}\label{thermal}
n_{\omega lm}= {\Gamma_{\omega lm}\over e^{\hbar\tilde\omega/T} - 1},
\end{equation}
where
\begin{equation} \label{temp}
{1\over T}: = \left({\partial S\over \partial M}\right)_J
\end{equation}
is the thermodynamical temperature of the black hole,
\begin{eqnarray} \nonumber
\hbar\tilde\omega := \hbar\omega - \Omega_H \hbar m, \\
\Omega_H := T\,\left({\partial S\over \partial J}\right)_M,  \label{def}
\end{eqnarray}
and $\Gamma_{\omega lm}$
is the absorption probability of the black hole in the mode $\omega lm$. 
To avoid confusion, let us note that the absorption probability is
different from the so-called absorption cross-section $\sigma_{abs}$
and is related to the later according to (\ref{cross-prob}).
Thus, with our
choice of the probability distribution, the black hole emission
spectrum is guaranteed to be thermal in the sense that it has the
form (\ref{thermal}), independently of details of the 
microscopic dynamics.
Moreover, as we show in the Appendix C, our choice of
the probability distribution guarantees the correct
($T^2$) dependence of the luminosity of a non-rotating
black hole on its temperature. This also holds
independently of the black hole microscopics.  

Thus, it might seem that all of the properties of the
black hole emission spectrum are correctly reproduced
as soon as one uses a correct statistical ensemble, and
that this holds no matter what is the microscopic
description.\footnote{%
It is interesting to note that traditional methods of statistical
mechanics can even be used to prove the validity of the 
generalized second law for the radiating black hole \cite{SecLaw}.
Thus, a lot of properties of black holes can be accounted for on
the basis of simple statistical mechanics by assuming an ``atomic''
internal structure. Somewhat surprisingly, this holds independently
of details of this structure.}
This conclusion, however, would not be correct. 
Recall that the ordinary, classical black hole is known to have a very
special dependence of the absorption cross-section on the quantum numbers
of the particle being absorbed. In particular, $\sigma_{abs}^{(l)}(\omega)$
for a Schwarzschild black hole is very small for large
angular momentum $\hbar l$, that is for 
$l/GM \gg \omega$. Also, for the absorption of the minimal coupled
scalar field, the absorption cross-section of
s-waves $\sigma_{abs}^{(0)}(\omega)$ is 
approximately constant and equal to the 
horizon area for all $\omega$ that matter in (\ref{thermal})
(see, for example, \cite{Gidd}). Also, the absorption 
cross-section $\sigma_{abs}(\omega)$ is a 
smooth function of the frequency. Only when these (and other)
properties are accounted for by a microscopic description 
can one be satisfied with this description. Thus, details 
of the microscopic dynamics become important
when one studies properties of the black hole absorption cross-section.
One could worry, for example, that
the quantum mechanical description of the black hole, although
giving a thermal emission spectrum in the above sense, does
not reproduce correctly the dependence of the absorption
coefficient on the quantum numbers $\omega, l$ etc. Also,
because the  space of states is discrete,
one could worry that the absorption properties of the
quantum black hole will be similar to those of an atom,
that is, that the absorption spectrum will consist of
distinct lines. This would correspond to 
the absorption coefficient very different from the one predicted
by the classical theory. 

In this paper we make a first step towards investigation of
the properties of the black hole absorption cross-section as predicted
by the non-perturbative quantum gravity approach.
More precisely, we study the emission and absorption of 
s-waves by a non-rotating, uncharged black hole. 
In order to be able to use the standard technology from 
statistical mechanics, we restrict our attention to a special 
set of lines in the emission spectrum. This allows us
to use an analogy between the quantum states of black holes
and the states of a gas of atoms. Although
our results concern the most uninteresting (however the
most dominant in the absorption) case of s-waves, the reader should bear in
mind that this paper is a first attempt to study the
emission-absorption properties of black holes within the
approach of non-perturbative quantum gravity. 

The paper is organized as follows. In Sec. \ref{sec:states} we
describe in some details the quantum states of black holes.
Sec. \ref{sec:specl} is 
devoted to the study of the spectrum. We study the
emission and absorption of s-waves by considering
a special set of lines in the emission spectrum.

\section{Quantum states}
\label{sec:states}

A description of 
quantum states is now available for various types of non-rotating black holes 
and it is expected that the main features will be carried over 
to the rotating case. Let us
first describe the quantum states of an uncharged, non-rotating black hole,
and then make some comments as to the rotating case.
(for details see \cite{K,ABCK}). Let $\B$ be the intersection
of the black hole horizon with a spatial hypersurface;
$\B$ has the topology of a 2-sphere. 
A state, in particular, is defined by a set of points 
$p_1,\ldots,p_n$ on 
$\B$ labelled by spins $j_p$. In what follows the points 
on $\B$ labelled by spins are referred to as punctures. 
Each set 
$$\P=\{j_{p_1},\ldots,j_{p_n}\}$$ 
of punctures gives rise to a Hilbert space ${\cal H}_\P$ 
of quantum states of the connection on $\B$, the connection
describing the effective black hole 
degrees of freedom. A black hole state is defined
by: (i) a set $\P$ of punctures on $\B$; (ii) 
a vector from the Hilbert space ${\cal H}_\P$.
The dimension of ${\cal H}_\P$, for a large
number of punctures, grows as
\begin{equation}\label{dim}
{\rm dim} {\cal H}_\P \sim \prod_{j_p\in\P} (2j_p+1)
\end{equation}

The quantum states described are eigenstates of the operator 
\cite{Area} measuring the area of $\B$.
Given a state defined by a set $\P$ of 
punctures the corresponding eigenvalue of the 
area operator is 
\begin{equation}
8\pi l_p^2 \gamma \sum_{j_p\in\P} \sqrt{j_p(j_p+1)},
\label{qarea}
\end{equation} 
where $l_p$ is the Planck length and $\gamma$ is the so-called
Immirzi parameter 
that arises due to a quantization ambiguity,
as is discussed 
in \cite{RT}. Given a black hole with horizon area $A$ one can 
introduce the statistical mechanical
entropy $S$ of the black hole, defined as the 
logarithm of the number of different quantum states that have an
area eigenvalue within the interval of width $l_p^2$ about $A$. One obtains
\begin{equation}
S = {\gamma_0\over 4 l_p^2 \gamma}\,A,
\label{entropy}
\end{equation}
where $\gamma_0$ is a numerical constant $\gamma_0\approx0.12$.
As we have mentioned above, the same value for the entropy can 
be obtained by considering the ``area'' canonical ensemble, defined
by the probability distribution (\ref{distr}) (see the Appendix A for
details). 

A detailed picture of quantum states 
for a rotating black hole is not yet available. 
Thus, we will not treat rotating black holes in the main body
of the paper. The only place where rotating black holes
are dealt with is the Appendix B, where the thermal 
character of the emission spectrum from a general
rotating charged black hole is discussed. For the
purposes of that discussion we shall make here a
minor assumption about the rotating black hole states.
Namely, we assume
that the rotating black hole can be described by quantum
states similar to those discussed above, i.e., 
that among the quantum states describing
the black hole, there are ones that are eigenstates of
both the area and the angular momentum operators. We assume
that the area spectrum is still given by (\ref{qarea}),
and denote the angular momentum eigenvalues by $J[\Gamma]$.
We will not need any further assumptions about these
quantum states.

\section{Spectrum}
\label{sec:specl}

We restrict our consideration to 
uncharged, non-rotating black holes, for which a complete microscopic
description of states is available, as was described above.
This description is used here to deduce some properties of
the absorption cross-section for such black holes.
Our results also provide a somewhat 
interesting picture of ``atoms'' of surface geometry. This section 
heavily uses the details of the microscopic description
of states.

To introduce our main physical idea let us heuristically think of punctures
on $\B$ as ``atoms'' of surface geometry. According to (\ref{qarea})
each atom gives a contribution to the area of $\B$ that depends on
its spin $j_p$. Thus, we say that atoms can reside 
in different quantum states, which are labelled by 
the quantum number $j_p$. Then, just as real atoms can
jump from one excited state to another emitting quanta of radiation,
our atoms can undergo transitions in which spin $j_p$ changes, thus
changing the horizon surface area of black hole and emitting radiation.

We assume that in the process of evolution described by an 
appropriate Hamiltonian  operator,
individual atoms jump from one 
quantum state to another, and the line intensities are given by 
Fermi's formula. We consider here
only emission processes involving just a single atom;
as we shall see, this simplifies the whole discussion
considerably. Note, however, that in this way we
will be able to analyze only a part of the emission
spectrum. To get the complete spectrum one should
consider processes involving simultaneously any
number of atoms.
We believe, however, that the results we obtain for this
simplified case give one an important insight on what can be
expected in the more general situation. Indeed, on general
grounds one might expect that transitions involving simultaneously
more than one puncture are much less probable and that, 
therefore, the transitions
we consider here determine the form of the spectrum.

The probability of a transition involving just a single atom 
is then given by the expression (\ref{fermi}), with $\hat{V}$ being 
a part of the quantum Hamiltonian responsible for the
transitions of this type. The intensity of a
line is given by (\ref{fermi-2}), with the labels $\Gamma,\Gamma'$
of the initial and final states being replaced by
initial and final spins $j,j'$ of the atom involved in the
transition. Since the final state
is $(2j'+1)$ degenerate, one should also multiply the 
intensity (\ref{fermi-2}) by
the degeneracy. 
Because we are using the analogy between a state of the black 
hole and a state of the gas of atoms, it is natural to
replace the probability $p(\Gamma)$ in (\ref{fermi}), 
(\ref{fermi-2}) by the number $n_j$ of atoms in the initial
state $\Gamma$ that are
in one and the same state $j$. The function $n_j(A)$, which gives
the number $n_j$ of ``atoms'' excited to the level $j$ among all the
``atoms'' that compose a black hole of given area $A$, is introduced, and
its properties are discussed in the Appendix A. This function
is analogous to the occupation number in the statistical
mechanical treatment of the harmonic oscilator. Similarly to
the case of the harmonic oscilator, the thermal character 
of radiation in our case comes from the properties of the 
function $n_j(A)$.

In the simplest case of the absorption of s-waves, the final state
of the field describing the radiation is spherically symmetric. Thus,
one can integrate out the angular
dependence in (\ref{fermi-2}). Let us denote the corresponding
integral of the squared absolute value of the 
transition matrix element by $\Phi_{jj'}$.
Let us also absorb in $\Phi$ all constant unimportant factors.
To get rid of the $\delta$-function
in (\ref{fermi-2}) one integrates over $d\omega$.
Thus, the intensity of the line $j\to j'$ becomes
\begin{eqnarray}
I(j\to j') & = & \omega_{jj'}^3\,n_j(A)\,\Phi_{jj'}\,(2j'+1),
\label{intensity} \\ 
\hbar\omega_{jj'} & = & \left( {dM\over dA} \right)\,(A_j-A_{j'}).
\label{rel-2}
\end{eqnarray}
Note that we have multiplied the whole expression by $2j'+1$,
which is the 
degeneracy of the final state. The latter formula is what one obtains
from (\ref{rel}) considering transitions that change only one puncture.
Here $A_j = 8\pi l_p^2 \gamma \sqrt{j(j+1)}$. 

The expression (\ref{intensity}) is the desired formula for the
emission line intensity.
Knowing the matrix elements of the transition Hamiltonian one 
could immediately calculate the intensities of all the lines
in the emission spectrum. 
It is interesting, however, that some information
about the overall form of the spectrum can be obtained even
without knowing the Hamiltonian. 

We are interested in getting the spectrum of a large black hole,
that is a black hole whose horizon area is large in Planck
units. As we discuss in the Appendix A, most of the area of a large
black hole is due to the contribution from atoms
whose spin $j=1/2$. This is, for example, manifested by the fact 
that $n_{1/2}(A)\to\infty$ as $A\to\infty$, while all other
occupation numbers $n_j$ remain finite.
An implication of this fact for the 
emission-absorption spectrum is that the lines corresponding
to transitions to and from $j=1/2$ are most intensive,
for a large black hole by far more intensive then any other
lines. For emission spectrum, for example, this would mean that
there is effectively only one intense emission line $1/2 \to 0$, and
all other lines are of negligible intensity as compared with this
one. Of course, such a spectrum is very far from thermal one. 
Note that this would not be in a contradiction with the general
result (\ref{thermal1}). Indeed, the spectrum in this case 
is still of the form (\ref{thermal1}), but the absorption 
coefficient is distributional and picked at a single frequency.

It turns out, however, that the line $1/2 \to 0$ is forbidden
by selection rules. And, as we shall see, with this line removed
from the spectrum, the remaining lines form a spectrum close to
thermal. The selection rule forbidding the line 
$1/2 \to 0$ is independent of the details of the Hamiltonian
describing the system. It arises because of the requirement 
of gauge invariance. As we have said above, a quantum state
of black hole is described by a collection of spins
labelling the punctures and by a state of Chern-Simons theory
on the punctured horizon surface. For state of Chern-Simons
theory to exist, one condition should be 
satisfied by the spins labelling the punctures. Namely,
the decomposition into irreducible representations 
of the tensor product of representations labelling the punctures 
should contain the trivial representation. In fact, the 
multiplicity of the trivial representation is exactly the 
dimension (\ref{dim})
of the Hilbert space of Chern-Simons theory. Translated on
the language of spins labelling the punctures, this 
condition states that the (algebraic) sum of all spins 
should be an integer. Only such configurations of
spins correspond to black hole states.
Let us now return to the transition $1/2 \to 0$. If the
initial state of the black hole is such that the sum 
of all spins gives an integer, then the state
obtained after this transition does not satisfy this
condition. Thus, the transition $1/2 \to 0$ should be forbidden.

Since the transitions 
from $j=1/2 \to j=0$ are forbidden, we have to deal only with with $j=1$ and
higher. For these values of spins $j$ the expression for $n_j(A)$
simplifies. Indeed, as we discuss in the Appendix A, large black
holes correspond to values of $\alpha$ very close to $\tilde\alpha$,
the later being the proportionality coefficient between the entropy
and the area. Thus, for $j>1/2$, the occupation number $n_j$ in the limit
of large $A$ becomes independent of the area and approximately 
equal to $n_j(\tilde\alpha)$. It is
not hard to find its asymptotic behavior for large $j$:
\begin{equation}\label{asympt}
n_j \propto j e^{-\tilde\alpha A_j}.
\end{equation}
Note that this is a function of $j$ only.

After these preliminaries, we can begin the discussion of properties
of the spectrum. To find the emission spectrum 
$I(\omega)$ one has to find the energy 
$I$ emitted per frequency
interval as a function of frequency $\omega$. 
Let us note that the relations (\ref{intensity}), (\ref{rel-2}) have the form
\begin{eqnarray}
I(j\to j') = \omega^3_{jj'} f_1(j,j')
\label{r1} \\
{\hbar\omega_{jj'}\over T} = f_2(j,j'),
\label{r2}
\end{eqnarray}
where $f_1(j,j')=n_j\Phi_{jj'}(2j'+1), f_2=\tilde\alpha(A_j-A_{j'})$ 
are functions of $j,j'$ only. Here $T$ is the
thermodynamical temperature of the black hole (\ref{temp}).
To find the total intensity $I(\omega)$ of the radiation emitted
as a function of frequency, one has to divide the
frequency range of interest into small intervals,
and sum the intensities of all lines that belong to each 
particular interval. Then the relations (\ref{r1}), (\ref{r2}) 
can be thought of as giving the parametric
dependence of $I$ on $\hbar\omega/T$ (the parameters
being $j,j'$). Thus, we find that 
$I(\omega)$ depends on the frequency $\omega$ only in the
combination $\hbar\omega/T$ 
$$I=I(\hbar\omega/T).$$

It is not hard to find the properties of this function in the
two limiting cases $\hbar\omega/T\to 0, \hbar\omega/T\to\infty$.
Indeed, the presence of $\omega^3$ in (\ref{r1}) guarantees
that $I\to 0$ as $\hbar\omega/T\to 0$. On the other hand, one can
use the asymptotics (\ref{asympt}) of $n_j$ to determine the
behavior of $I$ for large frequencies. Indeed, the
most intensive line with frequency $\omega$ (for large $\hbar\omega/T$)
is the one corresponding to the transition $j(\omega)\to 0$, where
$$\hbar\omega/T=\tilde\alpha A_{j(\omega)}.$$ 
The intensity of this line is proportional to 
$\exp{-\tilde\alpha A_{j(\omega)}}=\exp{-\hbar\omega/T}$.
The other, less intensive lines
with the same frequency correspond to transitions 
$j(\omega)+k\to k$; their intensity is 
$\exp{-\tilde\alpha A_k}$ less. Summing the intensities of
all these lines, one gets the asymptotics of the emission
intensity for large frequencies:
\begin{equation}
I(\omega) \propto
\exp{\left( -\,{\hbar\omega\over T} \right) },
\label{high}
\end{equation}
where $T$ is the thermodynamical temperature (\ref{temp}) of the black hole.
In other words, we find that the emission
spectrum $I(\omega)$ behaves at high frequencies as  
the Planck thermal spectrum of the temperature $T$. 

Of course, it is not surprising that we found some properties
of the thermal spectrum. Indeed, recall that the general
results of the Appendix guarantee that the intensity $I$ of the
radiation emitted in s-waves must have the form
\begin{equation}\label{4}
I(\omega)d\omega \sim 
{\omega^3 \sigma_{abs}^{(0)}(\hbar\omega/T) 
d\omega\over e^{\hbar\omega/T} - 1},
\end{equation}
where the absorption cross-section 
$\sigma_{abs}^{(0)}$ is a function of $\hbar\omega/T$.
Thus, the behavior of $I$ for large and small frequencies that
we have demonstrated above could have been derived
directly from the general result (\ref{4}). 
However, the analysis we just presented allows one
to learn more about the emission spectrum then just its
general form (\ref{4}). Note that (\ref{4}) still does not guarantee that
the emission spectrum is close to that of a black hole.
It can still happen that the absorption 
$\sigma_{abs}$ is very different from the one predicted by the classical 
theory. Indeed, from the classical theory we know that 
$\sigma_{abs}^{(0)}(\omega)$ 
is approximately constant for frequencies that matter in (\ref{4}). 
It could happen
that the absorption cross-section predicted by the quantum theory
is different.

To test the dependence of $\sigma_{abs}^{(0)}$ on $\omega$, 
we present a result of calculation
performed with formula (\ref{intensity}). We choose transition matrix
elements in the simplest possible way. Namely, we take $V_{jj'}$ to
be independent of $jj'$, and assume that no transitions to the
state $j=0$ are allowed, thus treating the state $j=1/2$ as
the `ground' state. In particular, this automatically takes
into account the selection rule that forbids the transitions
$j=1/2 \to j=0$. The corresponding spectrum is shown in FIG. 2.
The first plot shows lines with their intensities computed via 
(\ref{intensity}). Each line is shown as an impulse at a
point on the $\hbar\omega/T$ axis. The height of an impulse represents
the intensity of the line. We are interested in
line intensities only up to an overall multiplicative factor; thus, 
no units are shown on the $y$-axis. The second plot shows the total
intensity $I(\omega)$ as a function of frequency. To get
$I(\omega)$ here we divided the frequency range into
small intervals and added the intensities of all the
lines belonging to the same interval.
Although the emission occurs only at certain frequencies, the
enveloping curve of the spectrum is perfectly thermal, as 
is clear from the second plot. This means that, although the
absorption coefficient is distributional, its non-zero value
is approximately constant for the whole range of
frequencies, in agreement with the classical theory. 
Note that the total intensity
decreases exponentially as $\exp{(-\hbar\omega/T)}$ (second plot),
in accordance with our conclusion above.
The notable fine structure of emission lines (first plot) also
deserves attention. Thus, to the extent the naive model described is
correct, the properties of the absorption cross-section 
reproduce those of the classical black hole. It would be
much more interesting, however, to try to compute the
absorption cross-section of s-waves exactly and compare it with
the classical expression (the horizon area). We hope to 
perform a calculation to this effect in the future.

\begin{figure}[t]  
\centerline{\hbox{
\psfig{figure=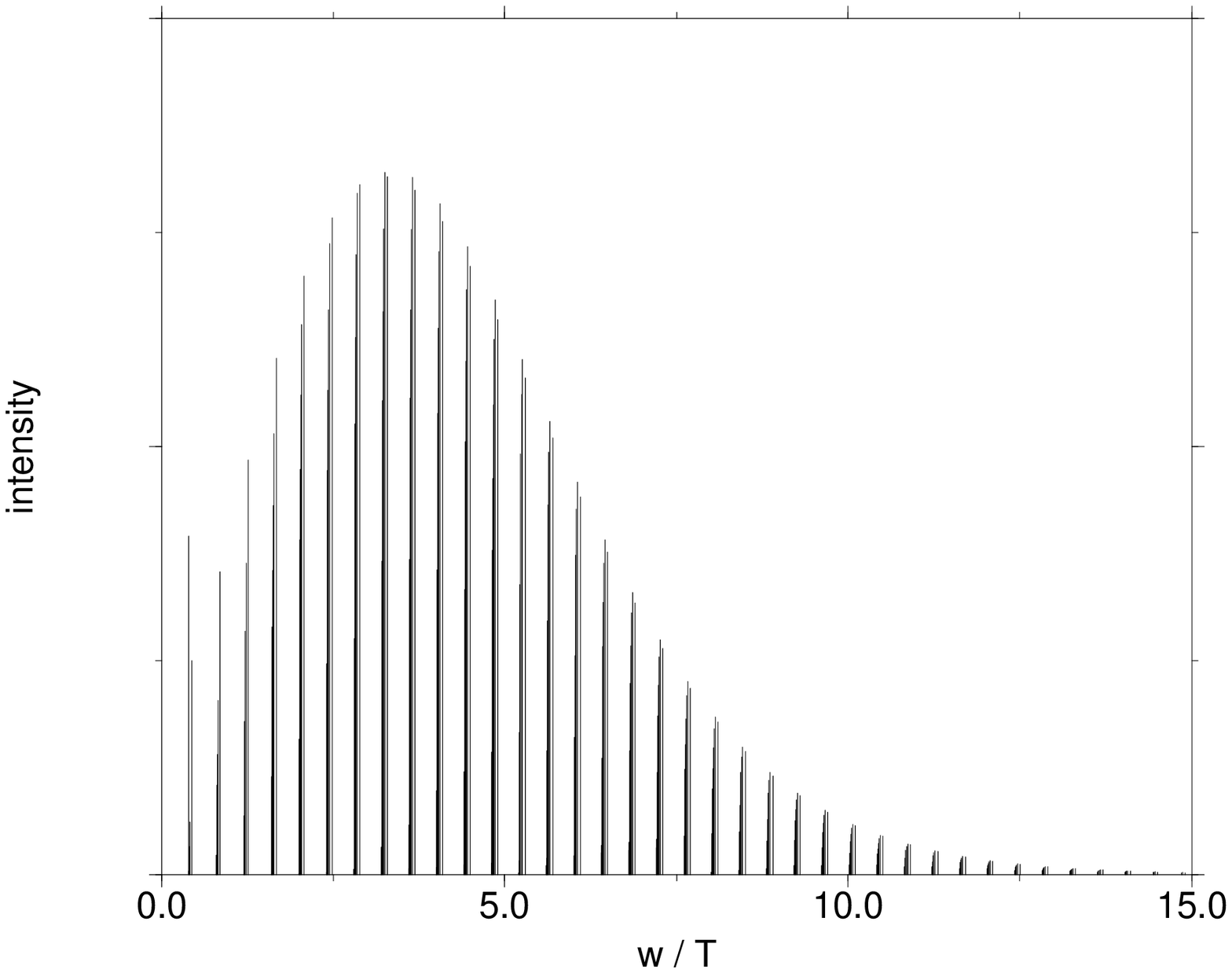,height=2in}}}
\centerline{\hbox{
\psfig{figure=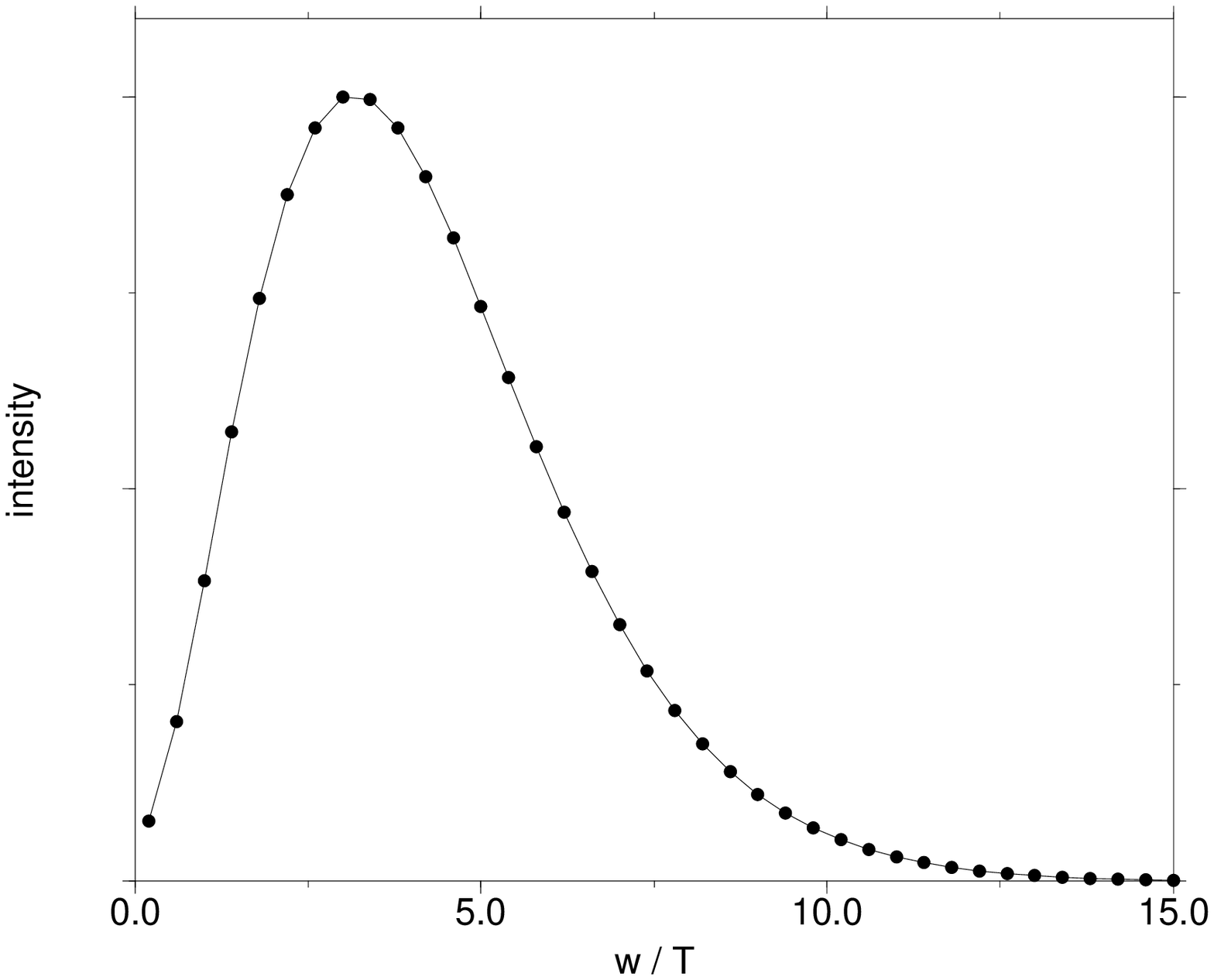,height=2in}}}
\caption{Emission lines with their intensities (above) 
and the total intensity of the radiation 
emitted per frequency interval as a function of frequency (below).}
\end{figure}

Several comments are in order. (i) It is instructive to compare
the results we have obtained with the results
of Bekenstein and Mukhanov \cite{BM}. Using a
simple model of quantum states of the black hole they
found that the emission spectrum consists of discrete lines with
the intensity of high frequency lines exponentially 
damped as in the thermal spectrum. Their model also predicted 
that the smallest frequency emission line is 
approximately the one with the largest intensity,
thus the spectrum being very different in form from
the thermal one. In contrast, the spectrum we have
found is much closer in form to the thermal spectrum.
(ii) Let us compare our results with the results
on the black hole emission-absorption spectra 
obtained using the microscopic description based
on D-branes \cite{Dbranes}. The two descriptions
are similar in the sense that both describe black
holes as composed of some elementary fundamental
building blocks: the flux lines of gravitational
field as ``atoms'' of geometry in our picture,
and D-branes in the string picture. In both
cases thermality is a result of averaging 
over a large number of microscopic states. Thus,
our approach gives the same answer to 
the question of information
loss as string theory. There are important
differences, however. Unlike the case of string
theory, our description is not
sypersymmetric and works entirely in four spacetime
dimensions. Because of this, the black holes
whose description is still problematic in string
theory, such as non-rotating, uncharged black holes
in four dimensions, are the ones that are
easiest to describe in our framework. Also,
our approach is more geometric, the black
hole degrees of freedom ``live'' in our approach
on the black hole horizon. Thus our approach 
directly accounts for the black hole entropy as
being proportional to the area. In string
theory description, it is harder to see ``where'' the degrees
of freedom of the black hole ``live'', and the 
area of the horizon plays no role, the black 
hole entropy being thought of as a function of
mass and other charges. Thus, the two approaches
are complimentary more than contradicting. 
We must admit, however, that the string description
of black holes gives results that precisely 
agree with the results that follow from semi-classical
considerations. The results of our approach are so
far more modest, the agreement with the
semi-classical results being reached only at
a qualitative level.
(iii) Although we have found that the spectrum is
discrete, it may be the case that the discrete
lines in the emission spectrum get smeared out by
quantum fluctuations, as it was argued for in \cite{Makela,Note}.
Thus, our results by themselves 
do {\it not} imply that the black hole spectrum in 
loop quantum gravity is discrete. Further work is necessary to
understand implications of the quantum fluctuations of the black hole horizon
for the emission spectrum.
(iv) One has to keep in mind that only the 
transitions involving one atom were considered.
It is known \cite{Carlo&,Carlo} that one obtains a quasi-continuous
spectrum considering transitions involving 
an arbitrary number of punctures. Thus, another possibility is that 
the discrete
spectrum we have found is only a part of a complete
quasi-continuous emission spectrum of the quantum black hole.
It might be, however, that
the transitions involving simultaneously many
atoms are highly suppressed (or forbidden) as
compared with the one-atom transitions. Thus, the
dynamics of the theory may be such that  
the lines analyzed in this paper are the most intensive
ones, determining the form of the emission spectrum. However, certainly
a more quantitative argument in support of this is necessary.

\acknowledgements

The author is indebted to A. Ashtekar for many enlightening discussions,
and to J.\ Baez and D.\ Marolf for suggestions on the first versions
of the manuscript. 
This work was supported in part by the NSF grant PHY95-14240 and by
the Eberly Research Fund of Penn State University. The author is also
grateful for the support received from the Erwin Schr\"odinger International
Institute for Mathematical Sciences, Vienna, where 
this work was partially done.

\appendix
\section{Area canonical ensemble}

In this appendix we introduce the `area' canonical ensemble \cite{GeomEntr}
and discuss some of its properties.
Let us introduce a parameter $\alpha$, which 
will be referred to as the intensive parameter conjugate to the
area, and consider the following function of $\alpha$, the
statistical sum
\begin{equation}
Q(\alpha) = \sum_\Gamma e^{-\alpha A[\Gamma]}.
\end{equation}
Thus, this is a Gibbs-type statistical ensemble; however, the area
is used instead of energy. Knowing the statistical sum $Q(\alpha)$
as a function of intensive parameter one can find all other
thermodynamical functions using the standard thermodynamical
relations. In particular, the dependence $S(A)$ of
the entropy on the area can be deduced from the dependence of
the expectation value of the area $A(\alpha)$ and
the entropy $S(\alpha)$ on the parameter $\alpha$:
\begin{eqnarray} \label{A}
A(\alpha) = \sum_\Gamma A[\Gamma]  p[\Gamma] =
- {d\ln{Q}\over d\alpha}, \\ \label{S}
S(\alpha) = \sum_\Gamma p[\Gamma] \ln{p[\Gamma]} = 
\alpha A(\alpha) + \ln{Q(\alpha)},
\end{eqnarray}
where 
\begin{equation}\label{p}
p[\Gamma]={1\over Q} e^{-\alpha A[\Gamma]}.
\end{equation}
This then gives a parametric dependence of $S$ on $A$.

In the particular case of the states $\Gamma$ given by those
described in the Sec. \ref{sec:states}, 
the statistical sum $Q(\alpha)$ can be easily
calculated:
\begin{equation}
Q(\alpha) = \prod_j {1\over 1-(2j+1)e^{-\alpha A_j}},
\end{equation}
where $A_j=8\pi l_p^2\gamma \sqrt{j(j+1)}$. The factors of 
$(2j+1)$ appear here because each ``atom'' state is $(2j+1)$ degenerate.
One can then
find the following expression for the mean value $A(\alpha)$
\begin{equation} \label{3}
A(\alpha) = \sum_{j=1/2}^\infty\,A_j\,n_j(\alpha),
\end{equation}
where $n_j(\alpha)$ is given by
\begin{equation}
n_j(\alpha) = {(2j+1)e^{-\alpha A_j}\over 1 - (2j+1)e^{-\alpha A_j}}.
\label{nparticles}
\end{equation}
This function plays an important role in our discussion of
the properties of the emission spectrum in Sec. \ref{sec:specl}.
It is easy to see that each function $n_j(\alpha)$ plays the role of the
mean number of punctures that have the spin $j$. According to
(\ref{A}), $\alpha$
controls the mean value of the black hole area. Thus, one can use $A$
instead of $\alpha$ as the argument of $n_j$. The function 
$n_j(\alpha(A))$ then,
given the black hole horizon area, tells how many there are atoms excited
to the level $j$ among all the atoms that compose the black hole. 

Some interesting properties of the ``area'' canonical 
ensemble can be derived from the properties of $n_j$. Note
that $n_j(\alpha)$ is finite for all $j$ for sufficiently large
$\alpha$. However, as $\alpha$ decreases, which corresponds
to the increase in the ``temperature'' in this ensemble,
for the value $\tilde\alpha=\ln{2}/A_{1/2}$ of $\alpha$ 
the denominator of $n_{1/2}(\alpha)$ becomes zero. Thus, as
$\alpha$ approaches the value $\tilde\alpha$, the occupation
number $n_{1/2}\to\infty$, with all other occupation numbers
remaining finite. Correspondingly, the physical quantities such as
the area and entropy diverge in the limit $\alpha\to\tilde\alpha$.
Formally, this corresponds to a phase transition in our
statistical ensemble. The reason for this
phase transition is the fact that density of states
grows linearly with the area. Indeed, it is not
hard to show that the phase transition of the type
described occurs in the ``area'' statistical ensemble
if and only if the density of states grows exponentially
with the area.
An analogous phase transition is known to occur
for ordinary statistical systems described by the Gibbs
ensemble, for which the density of states grows 
exponentially with the energy. Thus, the fact
that the physical quantities such as $A, S$ become
large as $\alpha$ approaches the critical value $\tilde\alpha$
imply the linear dependence of the entropy on the
area for large areas: $S=\tilde\alpha A$. 
The same expression for the entropy can be obtained by a
simple counting of states. This supports the validity of
the usage of the ``area'' canonical ensemble. Let us also note 
that, because $n_{1/2}\to\infty$ as $A\to\infty$, 
for large areas most of the ``atoms'' in the 
ensemble are in the state with $j=1/2$. This
last fact is important in our discussion of the
properties of the spectrum in Sec. \ref{sec:specl}.

Several comments are in order. (i) It is interesting to note
that, while the usual energy canonical ensemble is
not applicable to description of black holes, the ``area''
canonical ensemble is. We do not know whether there
is any deep physical significance behind this fact.
It might simply be that the ``area'' canonical ensemble
is a relevant description for systems whose density of
states grows linearly with the area. 
Note, however, that the probability distribution $p[\Gamma]$
given by the ``area'' canonical ensemble is also 
consistent with the properties of the black hole radiation.
Indeed, as is shown in the next Appendix, this probability
distribution guarantees the correct thermal properties
of the black hole emission spectrum. Thus, 
the ensemble in which ``energy squared'' (area) is used 
instead of energy, which is quite a
bizarre ensemble from the point of view of ordinary statistical mechanics,
appears to be very suited for a description of black holes.
(ii) Let us comment on a relation between the above
``area'' ensemble description of black holes and the
path integral description \cite{PathInt}. In the later
approach, the statistical sum of a gravity system is
found by summing (integrating) over Euclidean field configurations
weighted with $\exp{(-S)}$, where $S$ the Euclidean
action functional evaluated on the corresponding field
configuration. The field configurations that matter
most in this path integral are given by the solutions
of the classical field equations. In the context of asymptotically 
flat spacetimes, the ``bulk'' part
of the action vanishes on the solutions, and the
only contribution to the action comes from the boundary
term. When evaluated on field configurations 
corresponding to Euclidean black holes, the 
boundary term renders always the value $A/4l_p^2$,
where $A$ is the black hole horizon area. This then
results in the black hole entropy equal to $A/4l_p^2$.
It might seem, that our prescription (\ref{p})
for the black hole statistical ensemble is inconsistent
with the path integral approach, in which the statistical weight
of any black hole spacetime is also given by $\exp{-\alpha A}$,
but the parameter $\alpha$ is set to the constant
value $1/4l_p^2$ and not allowed to vary. Note,
however, that for large black holes, as we discussed
above, the value of $\alpha$ in the ``area'' canonical
ensemble is very close to the value $\tilde\alpha$.
In fact, $|\alpha-\tilde\alpha|\sim l_p^2/A$. But
the value $\tilde\alpha$ is just the proportionality
coefficient between the entropy and the area. Thus, for
large black holes, the ``area'' canonical ensemble
statistical distribution effectively reproduces the
one of the path integral approach. In other words,
for black holes that are large in Planck units,
there is no inconsistency between the two approaches.
This is the best one can expect: the two prescriptions
agree in their domain of applicability. Indeed,
the semiclassical path integral approach can only 
be trusted for large black holes, as well as the
statistical mechanical description based on the
``area'' ensemble is applicable only to sufficiently
complicated systems with a large number of internal states.

\section{Thermal character of the spectrum}

In this appendix we show that, if the probability 
distribution of quantum states of the black hole 
is given by (\ref{distr}), then the spontaneous
emission spectrum is thermal in the sense that it
has the form (\ref{thermal}). We use the standard
thermodynamical argument, which states that the
emission spectrum of a system described by the
canonical ensemble is thermal. However, in the
case of black holes, we cannot refer to that argument
directly because area is used instead of energy in 
(\ref{distr}). Our argument is somewhat similar to
the one given in \cite{B}.

Let us introduce 
the so-called Einstein coefficients. We denote 
by $A_{\omega lm}$ the coefficient of spontaneous emission
in the mode $\omega lm$. It is defined so
that the mean population $n_{\omega lm}$ of this mode due to
spontaneous emission is exactly $A_{\omega lm}$. We denote
by $\bdown_{\omega lm}$ the Einstein coefficient of stimulated
emission. It is defined in such a way that, if the number 
of incoming quanta in the mode $\omega lm$ is $k_{\omega lm}$,
then the population of this mode in the out-going
radiation due to stimulated emission is $\bdown_{\omega lm} k_{\omega lm}$. 
We denote by $\bup_{\omega lm}$ the Einstein coefficient of absorption. 
It is defined in such a way that, if the number 
of incoming quanta in the mode $\omega lm$ is $k_{\omega lm}$,
then the fraction $\bup_{\omega lm} k_{\omega lm}$ of
this mode is absorbed by the system. For brevity, we will
sometimes suppress the dependence of these coefficients on $\omega,l,m$.

We will first show that there exists a relation between the 
coefficients $\bup,\bdown$.
One can write the following expressions for the 
ratio of these coefficients
\begin{eqnarray} \label{1}
{\bdown\over\bup} = {\sum_{\Gamma,\Gamma'}
p(\Gamma') W_{\Gamma'\to\Gamma} \over 
\sum_{\Gamma,\Gamma'}
p(\Gamma) W_{\Gamma\to\Gamma'} },
\end{eqnarray}
where both sums are taken over $\Gamma,\Gamma'$
subject to the conditions $\omega_{\Gamma'\Gamma}=\omega >0,
\Delta J_{\Gamma'\Gamma}=\hbar m$, and where $p(\Gamma)$ is
the probability (\ref{distr}). Now note that
\begin{equation}
p(\Gamma) = e^{\tilde\alpha A[\Gamma'] - 
\tilde\alpha A[\Gamma]}\,p(\Gamma') = 
e^{{1\over T}(\hbar\omega_{\Gamma'\Gamma} - \Omega_H \Delta J_{\Gamma'\Gamma})}
\,p(\Gamma') = e^{\hbar\tilde\omega/T}\,p(\Gamma'),
\end{equation}
where the definitions (\ref{temp}), (\ref{def}) were used
to write the second and third identity. Thus, we have
\begin{equation} \label{rel1}
\bup = e^{\hbar\tilde\omega/T}\,\bdown,
\end{equation}
where we have used the fact that 
$W_{\Gamma'\to\Gamma}=W_{\Gamma\to\Gamma'}$.

We can now relate the coefficients $B$ with the absorption
probability $\Gamma$. The fraction of incoming radiation 
that is absorbed by the black hole
consists of two parts: the part that is absorbed minus 
the part emitted in the processes of stimulated emission.
Thus, we have for the absorption probability
\begin{equation}\label{rel2}
\Gamma = \bup - \bdown.
\end{equation}
Note that the absorption probability $\Gamma$ is different
from the so-called absorption cross-section $\sigma_{abs}$.
The two are, however, related according to:
\begin{equation} \label{cross-prob}
\sigma_{abs}(\omega) = \sum_l \sigma_{abs}^{(l)}(\omega)
{\pi\over\omega^2} \sum_{l=0}^\infty \sum_m
\Gamma_{\omega lm},
\end{equation}
where we have also introduced the so-called partial
absorption cross-sections $\sigma_{abs}^{(l)}$.

Let us now discuss a relation between the coefficients
$A,\bdown$, which are not independent. Let us assume
that the interaction of the black hole with the field
describing the radiation quanta is linear, as it is
the case, for example, for the interaction of an atom
with electromagnetic radiation. In this case one can
use the standard formulas for the matrix elements 
of the creation and annihilation operators to conclude 
that the probability
of emission of a quantum in a mode in which there are
already $k$ quanta is given by $k+1$ times the probability
of spontaneous emission in the same mode. This yields
\begin{equation}\label{rel3}
A = \bdown.
\end{equation}
Now, using the relations 
(\ref{rel1})-(\ref{rel3}), we can conclude:
\begin{equation} \label{thermal1}
n_{\omega lm} = A_{\omega lm} = {\Gamma_{\omega lm}
\over e^{\hbar\tilde\omega/T} - 1},
\end{equation}
which is the well-known expression for the black hole 
radiation spectrum.

The above argument is very 
general. To prove (\ref{thermal1}) we have used only the
assumption that the distribution of states is given by (\ref{distr}),
and the relation (\ref{rel3}) that holds for a large class of interaction
Hamiltonians. This means that the thermal character of radiation 
from the quantum black hole is independent on the details of
microscopic description as well as on the details of the quantum
dynamics. It depends only on the form of the probability distribution
describing the black hole. 

\section{Luminosity}

In this appendix we restrict our attention to uncharged
non-rotating black
holes, and study the dependence of the total luminosity of the quantum 
black hole on its temperature. For a classical black hole,
the emission is dominated by s-waves, for which the
absorption cross-section is approximately equal to the
horizon area (see, e.g. \cite{Page}). 
This gives the luminosity proportional to
the squared black hole temperature. However, we cannot refer to this
argument for the case of a quantum black hole, because no
such property of the absorption cross-section in the quantum theory
was established. Thus, we give a different, applicable to a quantum black
hole argument to the same effect.

The total luminosity is defined  
as amount of energy lost by the black hole per unit time.
We show that the dependence of this on the black hole temperature is 
\begin{equation} \label{lum}
{d M\over d t} \sim T^2.
\end{equation}
As in the previous subsection, we find that this result 
is very general, that is, it is independent on the
black hole microscopics. Also, as above, we consider only
the emission due to massless quanta.

For the rate with which the black hole looses its mass one has
\begin{equation} \label{lum1}
{d M\over d t} = \sum_{lm} \int_0^\infty {d\omega\over 2\pi} 
\hbar\omega {\Gamma_{\omega lm}
\over e^{\hbar\omega/T} - 1},
\end{equation}
where we used the fact that the black hole 
is non-rotating $\Omega_H=0$. Our key observation is that we get (\ref{lum})
given that $\Gamma_{\omega lm}$ depends on $\omega$ only in the
combination $\hbar\omega/T$. Indeed, in this case we can introduce
the new integration variable $\hbar\omega/T$. Then (\ref{lum1})
is given by $T^2$ times the expression that does not depend on
the temperature anymore. Assuming that the luminosity is finite, we
get (\ref{lum}). Note that this result will hold even if 
the absorption cross-section $\sigma$ is distributional, 
which we can expect for a black hole
with a discrete spectrum.

For a Schwarzschild black hole the absorption probability $\Gamma$ 
is given by a relation similar to (\ref{rel2}) 
\begin{equation}\label{2}
\Gamma = \bup - \bdown \sim \sum_{\Gamma,\Gamma'}
\left[p(\Gamma') W_{\Gamma'\to\Gamma} - 
p(\Gamma) W_{\Gamma\to\Gamma'}\right],
\end{equation}
where the sum is taken over $\Gamma,\Gamma': \omega_{\Gamma'\Gamma}=\omega$.
Let us note now that terms in the sum in (\ref{2}) depend solely on
the quantum numbers labelling $\Gamma,\Gamma'$: there is no dependence
on the temperature of the black hole. The only place where the dependence
on the temperature 
comes into play is the condition $\omega_{\Gamma'\Gamma}=\omega$.
To see this, let us rewrite this condition in terms of 
$\Delta A_{\Gamma\Gamma'}$. Using (\ref{rel}) we get
\begin{equation}
\hbar\omega = \hbar\omega_{\Gamma'\Gamma} = 
{dM\over dA}\,\Delta A_{\Gamma\Gamma'} = 
\tilde\alpha\,T\,\Delta A_{\Gamma\Gamma'},
\end{equation}
where we have used the definition of $T$ (\ref{temp}) and the fact
that $S=\tilde\alpha A$.
Thus, the condition on states $\Gamma,\Gamma'$ involves $\omega$
only in the combination $\omega/T$. This simple observation proves 
the property (\ref{lum}).


\begin{thebibliography}{99}
\bibitem{K} K.\ Krasnov, On statistical mechanics of 
Schwarzschild black holes, Gen. Rel. and Grav. {\bf 30}, 
No. 1, 53-68 (1998).

\bibitem{ABCK} A.\ Ashtekar, J.\ Baez, A.\ Corichi, K.\ Krasnov, 
Quantum geometry and black hole entropy,
Phys. Rev. Lett. {\bf 80}, No. 5, 904-907 (1998).

\bibitem{ACK} A.\ Ashtekar, A.\ Corichi, K.\ Krasnov, 
Black hole sector of the classical phase space, CGPG preprint.

\bibitem{ABK} A.\ Ashtekar, J.\ Baez, K.\ Krasnov, 
Quantum geometry of black hole horizons, CGPG preprint.

\bibitem{BM} J.\ Bekenstein, V.\ Mukhanov,
Spectroscopy of the quantum black hole,  
Phys.\ Lett. {\bf B360}, 7 (1995). 

\bibitem{Hawk} S.\ Hawking, Particle creation by black holes, 
Commun.\ Math.\ Phys.\ {\bf 43}, 212 (1975).

\bibitem{Makela} J.\ M\"akel\"a, Black hole spectrum: continuous or 
discrete?, available as gr-qc/9609001.

\bibitem{Note} K.\ Krasnov, Area spectrum in quantum gravity,
Class. Quant. Grav., in press.

\bibitem{GeomEntr} K.\ Krasnov, Geometrical entropy from
loop quantum gravity, Phys.\ Rev.\ {\bf D55} 3505 (1997).  

\bibitem{SecLaw} J.\ D.\ Bekenstein, Statistical black hole
thermodynamics, Phys.\ Rev.\ {\bf D12}, No. 10, 3077-3085 (1975);

S.\ W.\ Hawking, Black holes and thermodynamics, Phys.\ Rev.\ {\bf D13},
No. 2, 191-197 (1976).

\bibitem{Gidd} S.\ Giddings, Quantum mechanics of black holes,  
Lectures given at Summer School in High
Energy Physics and Cosmology, Trieste, Italy, 1994, 
Published in Trieste HEP Cosmology 1994: 530-574.

\bibitem{Area} C.\ Rovelli and L.\ Smolin, Discreteness of area 
and volume in quantum gravity, Nucl.\ Phys. {\bf B442},
593 (1995); Erratum: Nucl.\ Phys. {\bf B456}, 734 (1995).

R.\ De Pietri and C.\ Rovelli, Geometry eigenvalues and scalar product 
from recoupling theory in loop quantum gravity, Phys.\ Rev. {\bf D54},
2664 (1996).
 
A.\ Ashtekar, J.\ Lewandowski, Quantum theory of geometry I: Area
operators, Class.\ Quant.\ Grav. {\bf 14}, 55 (1997). 

K.\ Krasnov, On the constant that fixes the area spectrum in 
canonical quantum gravity, Class.\ Quant.\ Grav. {\bf 15}, 
L1-L4 (1998).

\bibitem{RT} G.\ Immirzi, Quantum gravity and Regge calculus,
Nucl.\ Phys.\ Proc.\ Suppl. {\bf 57} 65-72 (1997).

C.\ Rovelli and T.\ Thiemann, The Immirzi parameter in quantum general
relativity, Phys.\ Rev. {\bf D57} 1009-1014 (1998).
 
\bibitem{Dbranes} S.\ R.\ Das and S.\ D.\ Mathur, Comparing
decay rates for black holes and D-branes, Nucl.\ Phys.\ {\bf B478},
561-576 (1996);

J.\ Maldacena and A.\ Strominger,
Black hole greybody factors and D-brane spectroscopy,
Phys.\ Rev.\ {\bf D55}, 861-870 (1997). 

\bibitem{Carlo&}M.\ Barreira, M.\ Carfora, C.\ Rovelli, Physics with 
non-perturbative quantum gravity: radiation from a quantum black hole,
Gen.\ Rel.\ and Grav.\ {\bf 28}, 1293-1299 (1996).
 
\bibitem{Carlo} C.\ Rovelli,  Loop quantum gravity and black hole physics,
Helv.\ Phys.\ Acta., {\bf 69}, 582 (1996).

\bibitem{PathInt} G.\ Gibbons and S.\ Hawking, Action integrals
and partition functions in quantum gravity, Phys.\ Rev.\ {\bf D15},
2752-2756 (1977);
 
J.\ D.\ Brown and J.\ W.\ York, Microcanonical functional
integral for the gravitional field, Phys.\ Rev.\ {\bf D47},
No. 4, 1420-1431 (1993). 

\bibitem{B} J.\ D.\ Bekenstein and A.\ Meisels, Einstein A and B  
coefficients for a black hole, Phys.\ Rev.\ {\bf D15}, 2775-2781
(1977).

\bibitem{Page} D.\ Page, Particle emission rates from a black hole:
Massless particles from an uncharged, non-rotating black hole,
Phys.\ Rev.\ {\bf D13}, No. 2, 198 (1976).

\end{thebibliography}
\end{document}